\documentclass[preprint,12pt]{elsarticle}
\usepackage{CJK,lineno}
\usepackage{amssymb}
\usepackage[pdftex,colorlinks]{hyperref}
\usepackage{multirow}
\usepackage[tbtags]{amsmath}

\biboptions{compress}

\begin{document}

\begin{frontmatter}

\title{Quantum differential cryptanalysis to the block ciphers}
\author{Hong-Wei Li$^{1,2,3,4}$}
\author{Li Yang$^{1,3}$\corref{1}}%\ead{yangli@iie.ac.cn}
\cortext[1]{Corresponding author email: yangli@iie.ac.cn}
\address{1.State Key Laboratory of Information Security, Institute of Information Engineering, Chinese Academy of Sciences, Beijing 100093, China\\
2.School of Mathematics and Statistics, Henan Institute of Education, Zhengzhou,450046,Henan, China\\
3.Data Assurance and Communication Security Research Center, Chinese Academy of Sciences, Beijing 100093, China\\
4.University of Chinese Academy of Sciences, Beijing 100049, China}

\begin{abstract}
Differential cryptanalysis is one of the most popular methods in attacking block ciphers. However, there still some limitations in traditional differential cryptanalysis. On the other hand, researches of quantum algorithms have made great progress nowadays. This paper proposes two methods to apply quantum algorithms in differential cryptanalysis, and analysis their efficiencies and success probabilities. One method is using quantum algorithm in the high probability differential finding period for every S-Box. The second method is taking the encryption as a whole, using quantum algorithm in this process.

\end{abstract}
\end{frontmatter}

\section{Introduction}
\noindent
Differential cryptanalysis plays a central role in attacking modern crypto systems, especially in block ciphers \cite{BS91}. Now, this method has been developed to various forms, such as truncated differential attack \cite{K94} and impossible differential attack \cite{B11}. However, current ciphers (such as AES)  were designed along the wide trail strategy to resist differential cryptanalysis. On the other hand, quantum computation based on quantum mechanics has been built up, and has shown great speedups over classical computation in some areas. It is thus conceivable to use quantum algorithms in differential cryptanalysis.

Deutsch and Jozsa \cite{DJ92} presented a quantum algorithm to distinguish a balanced Boolean function from a constant function efficiently without error, which first show exponential speedup over classical algorithm. Using the same network as the above algorithm, Bernstein and Vazirani \cite{BV93} gave a quantum algorithm to identity linear functions. Later, Simon \cite{DR97} suggested a quantum algorithm for finding the period of a Boolean function. Inspired by Simon's algorithm, Shor \cite{PW97} discovered polynomial-time algorithms for factoring integers and solving discrete logarithms. Different from the above algorithms which rely on some promises of the problems, Grover's algorithm \cite{G97} searches a target element in an unsorted database and shows a quadratic speedup over the classical one.

In recent years, researches of quantum algorithm mainly focus on developments of the above mentioned algorithms. For example, there are quantum tests for whether a function has some properties or $\epsilon$-far from it \cite{AR09,SEAR10,ME11}, and there are also quantum algorithms for learning of Boolean functions \cite{AR09,DEM13}, but still with a promise that the Boolean functions belong to a small special set. Meanwhile, there are quantum polynomial algorithms to approximate some problems \cite{DVZ06,YYH08,LY14}. Amongst these algorithms, \cite{LY14} gave an efficient algorithm to find some high probability differentials of a Boolean function. In \cite{MR15}, the authors gave quantum related-key attacks based on Simon's algorithm.

\paragraph{\bf Our contributions.}
Inspired by \cite{LY14,MR15}, using the result in \cite{LY14}, and combining with the classical differential cryptanalysis approach, we investigated the differential cryptanalysis based on quantum algorithm and gave quantum algorithms to implement the differential cryptanalysis.

\paragraph{\bf In contrast to previous works.}
In \cite{SHW14}, the authors gave properties of an S-box and proposed a classical automatic approach to find (related-key) differential characteristics. Regarding the quantum differential cryptanalysis methods, one must mention \cite{ZLZS15}, which presented a quantum differential cryptanalysis based on the quantum counting and searching algorithms, and obtained a quadratic speedup over classical one. Their quantum algorithm is used after the time that the differential characteristics has been found. Contrary to the above works, our quantum algorithms are to find the differential characteristics.

\section{Preliminaries}
\noindent
In this section, we give some preliminaries and notations, which will be used in the following sections.

Let $F:\,\{0,\,1\}^m\rightarrow \{0,\,1\}^n$ be a multi-output Boolean function with input $x=(x_1,\,x_2,\cdots x_m)$ and output $y=(y_1,\,y_2,\cdots y_n)$, where $m,\:n$ are both positive integers. Let $F(x')=y'$ and $F(x'')=y''$,
then $\triangle x=x'\oplus x''$ and $\triangle y=y'\oplus y''$ are called the input difference and output difference, respectively, where $\oplus$ is the bit-wise exclusive-OR. Hence, $$\triangle x=(\triangle x_1,\,\triangle x_2,\cdots \triangle x_m),$$ and $$\triangle y=(\triangle y_1,\,\triangle y_2,\cdots \triangle y_n),$$ where $\triangle x_i=x_i'\oplus x_i''$ and $\triangle y_i=y_i'\oplus y_i''$. The pair $(\triangle x,\,\triangle y)$ is called a \emph{differential}.

A \emph{differential characteristic} is composed of input and output differences, where the input difference to one round is determined by the output difference of the last round.

\subsection{Classical differential cryptanalysis}
\noindent
Differential cryptanalysis is a chosen-plaintext attack. It is usually used to attack various block ciphers. Roughly speaking, differential cryptanalysis is composed by two procedures:

1. Find some high probability differential characteristics.

2. According to the differential characteristics which have been found, test possible candidate subkey, then recover the key of the cryptosystem.

In this paper, our quantum algorithm is used at the first process. While in \cite{ZLZS15}, their quantum algorithm was at the second stage.

\subsection{The Bernstein--Vazirani algorithm}
\noindent
Before showing the Bernstein--Vazirani algorithm, we first give the following definition:

\noindent{\bf Definition 1}\quad
For a Boolean function $f:\,\{0,\,1\}^m\rightarrow \{0,\,1\}$, the Walsh transform of $f$ is
\begin{equation}\label{eq:a}
S_f(w)=\frac{1}{2^m}\sum_{x\in F^m_2}(-1)^{f(x)+w\cdot x}
\end{equation}
for all $w\in F^m_2$.

\noindent{\bf Definition 2}\quad
For a Boolean function $f:\,\{0,\,1\}^m\rightarrow \{0,\,1\}$, define the transform
\begin{equation}\label{eq:b}
U_f|x\rangle|y\rangle=|x\rangle|y+f(x)\rangle.
\end{equation}

note that $U_f$ is unitary.

Now let us illustrate the Bernstein--Vazirani algorithm.

1.\,Input the initial state $|\psi_0\rangle=|0\rangle^{\otimes m}|1\rangle$, then do the Hadamard transform $H^{\otimes (m+1)}$, the result is
\begin{equation}\label{eq:c}
|\psi_1\rangle=\sum_{x\in F^m_2}\frac{|x\rangle}{\sqrt{2^m}}
\cdot\frac{|0\rangle-|1\rangle}{\sqrt{2}}.
\end{equation}

2. Evaluate $f$ by using $U_f$, giving
\begin{equation}\label{eq:d}
|\psi_2\rangle=\sum_{x\in F^m_2}
\frac{(-1)^{f(x)}|x\rangle}{\sqrt{2^m}}\cdot
\frac{|0\rangle-|1\rangle}{\sqrt{2}}.
\end{equation}

3. Execute the Hadamard transform $H^{\otimes (m)}$ on the first qubit of $|\psi_2\rangle$, we have
\begin{equation}\label{eq:e}
\begin{split}
|\psi_3\rangle
&=\sum_{y\in F^m_2}\frac{1}{2^m}\sum_{x\in F^n_2}(-1)^{f(x)+y\cdot x}|y\rangle
\cdot\frac{|0\rangle-|1\rangle}{\sqrt{2}}\\
&=\sum_{y\in F^m_2}S_f(y)|y\rangle
\cdot\frac{|0\rangle-|1\rangle}{\sqrt{2}}.
\end{split}
\end{equation}
If we measure the first $m$ qubit in the computational basis, we will obtain $y$ with probability $S_f^2(y)$.

\subsection{Results after running the Bernstein--Vazirani algorithm}
\noindent
In this section, we show that we will get some high probability differentials after running the Bernstein--Vazirani algorithm several times.

\noindent{\bf Theorem 1}\quad \cite{LY14}
For a Boolean function $f:\,\{0,\,1\}^m\rightarrow \{0,\,1\}$, let $p=p(m)$ be a polynomial of $m$. Assuming one has run the Bernstein--Vazirani algorithm $p$ times, and has obtained a set $S$. Solving the linear systems of equations $S\cdot X=0$ and $S\cdot X=1$ respectively gives two sets $A^0$ and $A^1$. Then $\forall a\in A^i(i=0,1)$, $\forall \epsilon$, $0<\epsilon<1$,
\begin{equation}\label{eq:f}
\text{Pr}\left(1-\frac{|\{x \in F^m_2|f(x\oplus a)+f(x)=i\}|}
{2^m}<\epsilon\right)>1-e^{-2p\epsilon^2},
\end{equation}
where $\text{Pr}(E)$ denotes the probability of the event $E$ happens.

\section{Quantum algorithm to execute differential cryptanalysis}
\noindent
Assume the plaintexts and the ciphertexts of the block cipher we would attack are of length $k=lm$, and every S-box is a map $F$ from $\{0,\,1\}^m$ to $\{0,\,1\}^n$, where $m,\:n,\:l$ are all positive integers. In the following we give two technics to implement quantum differential cryptanalysis.

\subsection{The first method}
\noindent
For every S-Box $F$, let $F=(f_1,\,\ldots,f_n)$, where each $f_j\,(j=1,\,\ldots,n)$ is a Boolean function $\{0,\,1\}^m$ to $\{0,\,1\}$. For every $f_j$, run the Bernstein--Vazirani algorithm $p=p(m)$ times, and later solve a linear system of equations to get $A^0_j$ and $A^1_j$. If there exists $a\in A_1^{i_1}\bigcap A_2^{i_2}\bigcap \ldots \bigcap A_n^{i_n}$, where $i_j \in \{0,\,1\}, \:j=1,\,2,\,\ldots,\,n$, then $(a,\,i_1i_2\ldots i_n)$ is a high probability differential.
\begin{bframe}
\textbf{Algorithm 1.}

\textbf{Input:} An S-Box $F=(f_1,\,\ldots,f_n)$.

\textbf{Output:} Some high probability differentials of each $f_j\,(j=1,\,2,\,\ldots,\,n)$.

1 Let $\mathcal{H}$ := $\emptyset$, $\mathcal{A}$ := $\emptyset$, where $\emptyset$ is the empty set.

2 \textbf{for} $j=1,\,2,\,\ldots,\,n$ \textbf{do}

3\qquad \textbf{for} $p=1,\,2,\,\ldots,\,p(m)$ \textbf{do}

4\qquad\quad Run the Bernstein--Vazirani algorithm, and get an n-bit output $w$;

5\qquad\quad Let $\mathcal{H}$ := $\mathcal{H}\cup \{w\}$

\qquad\: \textbf{end}

6\qquad Solve the equations $\mathcal{H}X=0$ and $\mathcal{H}X=1$ to get $A_j^0$ and $A_j^1$, respectively.

7\qquad Output $A_j^0$ and $A_j^1$.

\quad \textbf{end}

\end{bframe}
\hspace*{6.5cm}

After running the Algorithm 1, we obtain $A_j^i\,(j=1,\,2,\,\ldots,\,n;\,i=0,1)$. In the following, we will analyse these sets to get some high probability differentials of a S-Box $F$.

We may choose first the $p(m)=cm$ (where $c$ is a constant and $c\geq 2$) in Algorithm 1, since this can make every vector $a$ in $A_j^i\,(j=1,\,2,\,\ldots,\,n;\,i=0,1)$ satisfy
$$\frac{|\{x \in F^m_2|f_j(x\oplus a)+f_j(x)=i\}|}{2^m}>\frac{1}{2}$$
with high probability according to \cite{LY14}.

In other words, for any vector $a$ in $A_j^i\,(j=1,\,2,\,\ldots,\,n;\,i=0,1)$, $(a,\,i)$ is a differential of $f_j$ with the probability more than uniform distribution.

If most of the $A_j^i\,(j=1,\,2,\,\ldots,\,n;\,i=0,1)$ have a great deal of vectors (for example, a half of the whole), then we will choose $p(m)$ to be more large (for example, $p(m)=m^2$). The purpose of doing this is to prevent $|A_j^i|$ (where $|A|$ denotes the cardinality of a set $A$) from being too large.

Otherwise we execute the following algorithm to find some high probability differentials of $F$.

\begin{bframe}
\textbf{Algorithm 2.}

\textbf{Input:} $A_j^i\,(j=1,\,2,\,\ldots,\,n;\,i=0,\,1)$.

\textbf{Output:} Some high probability differentials of $F$.

1 \textbf{for} each $a\in A_1^{i_1} \,(i_1=0,\,1)$ \textbf{do}

2 \qquad \textbf{for} $j=2,\,\ldots,\,n$ \textbf{do}

3 \qquad\qquad \textbf{for} $i_j=0,1$ \textbf{do}

4 \qquad\qquad\qquad \textbf{if} $a\in A_j^{i_j}$ \textbf{then}

 \qquad\qquad\qquad\qquad  $(x_a,\,y_a)\,:=\,(a,\,i_1\,\ldots i_j)$

  \qquad\qquad\qquad\: \textbf{end}

 \qquad\qquad\:\, \textbf{end}

5 \qquad\qquad \textbf{else if} $a\notin A_j^{0}$ and $a\notin A_j^{1}$ \textbf{then}

 \qquad\qquad\quad
 $(x_a,\,y_a)\,:=\,(0,\,0)$

 \qquad\qquad\quad \textbf{goto} 6

 \qquad\:\, \textbf{end}

6 \qquad Output $(x_a,\,y_a)$

 \quad \textbf{end}

\end{bframe}
\hspace*{6.5cm}

The outputs of Algorithm 2 will be some vectors like $(a,\,i_1\,\ldots i_n)$ or $(0,\,0)$. Those non-zero vectors are the high probability differentials that we are looking for, which will be used to construct differential characteristics. For convenience, let $\mathcal{A}$ be the set of these non-zero vectors.

Next, complete the remaining works just as the classical differential cryptanalysis do.

\paragraph{\bf Analysis of the first method.}

\noindent
Now, let us see the efficiency of the first method.

In Algorithm 1, the time of running the Bernstein--Vazirani algorithm (in order to evaluate the function $F$) is $np(m)$, and the time needed to solve the system of linear equations is $nq(m)$ (where $q(m)$ is another polynomial of $m$). So the total time of Algorithm 1 is $np(m)+nq(m)$.

The maximum time of running the Algorithm 2 is $O(2^n)$. In fact, this upper bound may be a little rough, because for some $a\in A_1^{i_1} \,(i_1=0,1)$, they may be not in $A_j^{0}$ and $A_j^{1}$, where the $j$ is much less than $n$.

Next, let us consider the success probability of the first method.

The vectors $(a,\,i_1\,\ldots i_n)\in\mathcal{A}$ obtained by Algorithm 2 all satisfy the inequality \eqref{eq:f} for every $i_j$ and corresponding $f_j\,(j=1,\,2,\,\ldots,\,n)$. The number of $x$ satisfying
\begin{equation}\label{eq:g}
\frac{|\{x \in F^m_2|f_j(x\oplus a)+f_j(x)=i_j\}|}{2^m}=1-\epsilon
\end{equation}
for two different $j=j_1$ and $j=j_2$ is at least $2(1-\epsilon)-1=1-2\epsilon$. From \eqref{eq:f} and \eqref{eq:g}, we can know that
\begin{equation}\label{eq:h}
\text{Pr}\left(\frac{|\{x \in F^m_2|F(x\oplus a)+F(x)=i_1\,\ldots i_n\}|}
{2^m}>1-n\epsilon\right)>(1-e^{-2p\epsilon^2})^n.
\end{equation}
From the above inequality \eqref{eq:h}, we see that if $\epsilon=\frac{1}{c_1n}$ (where $c_1\geq 2$ is a constant), $p=\frac{c_2}{\epsilon^2}=c_2c_1^2n^2$ (where $c_2\geq 1+\frac{\ln n}{2}$ is also a constant), then
\begin{equation}\label{eq:i}
(1-e^{-2p\epsilon^2})^n\geq (1-e^{-2c_2})^n\geq 1-ne^{-2c_2}\geq 1-\frac{1}{e^2}
\end{equation}

\textit{In summary}, let $p=\max\{p(m),\,c_2c_1^2n^2\}$, after a total time of $np+nq(m)+O(2^n)$, we will get a set $\mathcal{A}$ constituted by vectors like $(a,\,i_1\,\ldots i_n)$, which satisfy
\begin{equation}\label{eq:j}
\text{Pr}\left(\frac{|\{x \in F^m_2|F(x\oplus a)+F(x)=i_1\,\ldots i_n\}|}
{2^m}>1-\frac{1}{c_1}\right)>1-\frac{1}{e^2}.
\end{equation}

As compared to the above quantum algorithm, the classical algorithm need $2^{m+n}$ times computation to give the difference distribution table, from which one can easily know some high probability differentials. Generally speaking, the S-Box used in a block cipher is not large, i.e., $m$ and $n$ are both small, so $2^{m+n}$ is very small too. In other words, evaluation  of the difference distribution table is very efficient, our quantum algorithm does not show much speedup over the classical algorithm. However, that provide a new approach to the problem, and may throw light on some other questions.

The above method only focuses on each S-Box. In the following, we will give another method. The difference is it will focus on the entire process of the encryption.
\subsection{The second method}
\noindent
Recall that the difficulty in the differential cryptanalysis is to construct high probability differential characteristics. And in the classical differential cryptanalysis, high probability differential characteristics are unambiguously given, from which S-Box to which S-Box. In fact, the purpose of doing that is to find which input differences will probably lead to which output differences. In the following, we will give a quantum algorithm to complete this.

Assume $G:\,\{0,\,1\}^k\rightarrow \{0,\,1\}^k$ is a function which maps the plaintext $x$ to the input $y$ of the last round under a secret key $K$. Certainly, $G$ can be written as $G=(g_1,\,g_2,\ldots,g_k)$. Assume also there is a polynomial-size quantum circuit to evaluate $G$.

\paragraph{\bf The Method 2} will be composed of the following procedures.

\paragraph{\bf At first,} run an algorithm similar to Algorithm 1. Nevertheless, the input to the algorithm is $G$ instead of $F$, the outputs are some high probability differentials $B_j^0$ and $B_j^1$ of each $g_j\,(j=1,\,2,\,\ldots,\,k)$.

\paragraph{\bf Secondly,} operate an algorithm similar to Algorithm 2. The differences are the inputs, the procedures and the outputs. The inputs are $B_j^0$ and $B_j^1$ $(j=1,\,2,\,\ldots,\,k)$. The procedures do not include line 5. The outputs will be some high probability differentials $\mathcal{B}=\{(b,\,i_{j_1}\cdots i_{j_t})\}$, where $j_1,\,\cdots,\,j_t \in \{1,\,2,\,\ldots,\,k\}$ and $j_1<\cdots< j_t$.

The reason why we delete line 5 is that the purpose of Algorithm 2 is to find out some shared differentials of all $f_j\,(j=1,\,2,\,\ldots,\,n)$. If $a\notin A_j^{0}$ and $a\notin A_j^{1}$ for a $j$, then $a$ must not their shared differential. At this time, breaking out of the loop is for saving time. what we do in the second method is to find  some differentials of part $g_j\,(j=1,\,2,\,\ldots,\,k)$ sharing.

\paragraph{\bf Thirdly,} determine the subkey in the last round according to the differentials obtained.

\paragraph{\bf Analysis of the second method.} Let us consider the time complexity. In the first and second procedures, the running time are all polynomial of $k$. The time of the last procedure is determined by the high probability differentials we have obtained. If the probabilities of the differentials are very high, this method would probably succeed by using much less time. The superiority of this approach is that it avoid finding concrete high differential characteristics.
\section{Discussions and Conclusions}
\noindent
Because high probability differential characteristics are independent of the subkey of every round, we can construct an efficient quantum circuit to find some of them. This paper proposes two methods for applying quantum algorithms to differential cryptanalysis. Although the first method does not show much speedup over classical method because the total number of the differences of an S-Box is not very large in practice, and the analysis of the second method is not very elaborate, these two methods give us a new clue to resolute the problem. Maybe they can be used in some ciphers and show much more speedups over classical approaches.

\subsubsection*{Acknowledgments.}
This work was supported by  the National Natural Science Foundation of China under Grant No.61173157.

\end{document}